\documentclass[showpacs,twocolumn,epsfig,prl]{revtex4-1}
\usepackage{graphicx}
\usepackage{amssymb}
\usepackage{color,soul}
\usepackage{amsmath}

\newcommand{\ket}[1]{|#1\rangle}
\newcommand{\bra}[1]{\langle#1|}

\newcommand{\tr}{\mathrm{Tr}}

\begin{document}

\draft
\title{A Fluctuation Theorem for Arbitrary Quantum Bipartite Systems}

\author{Jung Jun Park$^{1}$}
%\email{hyok07@gmail.com}
\author{Sang Wook Kim$^{2}$}
\author{Vlatko Vedral$^{1,3}$}
\affiliation{Centre for Quantum Technologies, National University of Singapore, 3 Science Drive 2, 117543, Singapore$^1$}
\affiliation{Department of Physics Education, Pusan National University, Busan 609-735, South Korea$^2$}
\affiliation{Atomic and Laser Physics, Clarendon Laboratory, University of Oxford, Parks Road, Oxford OX13PU, United Kingdom$^3$}

\date{\today}
%Setup
%Hamiltonian
%Main result
%Applications

\begin{abstract}
\label{*abstract*}
We present a fluctuation theorem for quantum bipartite systems in which the subsystems exchange information with each other. Our information fluctuation theorem includes correlations by introducing a quantum mechanical mutual information content and a statistical probability distribution which allows one to consider statistical averages when computing the thermodynamic and informational quantities.  The fundamental limit of heat transferred from a heat reservoir for quantum bipartite systems involving the quantum mutual information is then derived. We illustrate the use of our thermodynamic inequality in quantum computing and discuss our results in terms of Landauer's principle.\end{abstract}
\pacs{03.67.-a,05.30.-d,89.70.Cf,05.70.-a}
% 05.70.Ln Nonequilibrium and irreversible thermodynamics
% 03.67.-a Quantum information
% 05.30.-d Quantum statistical mechanics
% 89.70.Cf Entropy and other measures of information
% 03.65.Ta Foundations of quantum mechanics; measurement theory
% 05.70.-a Thermodynamics

\maketitle

Generalizations of fluctuation relations to different physical scenarios have attracted great interest \cite{DE93, GG95, CJ97, GC99, HT00, JK00, US05, GC08, ME09, MC10, MC11, VV12, TS10, MM11, TS12, KF13, SJ15}. Recently, the fluctuation theorems describing bipartite systems have been studied in both classical and quantum domains \cite{TS10, MM11, TS12, KF13, SJ15}. In particular, studies have shown that the fluctuation theorems are described by work fluctuations with feedback controls based on an information exchange between a system and a memory \cite{TS10, TS12, MM11}, or heat fluctuations with an information exchange between two subsystems \cite{SJ15}, or entropy production with an information exchange between two subsystems \cite{TS12, KF13}.

In the fluctuation theorems for quantum bipartite systems, however, arbitrary initial states have not been considered. The previous works have assumed that the initial state starts in a product state or a classically correlated state. Moreover, the statistics of the information-thermodynamic quantities in the fluctuation theorems are typically obtained by performing a measurement on the subsystems at the initial and the final times, during the standard two-point measurement protocol (TMP) \cite{HT00, JK00, ME09, MC11, VV12}. For arbitrary quantum bipartite states, the resulting information fluctuation theorems with the TMP do not exhibit the quantum mechanical features of correlations between the subsystems because the initial measurement destroys correlations.

In this Letter, we show how the fluctuation theorem for arbitrary bipartite systems can be generalized to involve all the quantum mechanical features. To this end, we consider statistics of the information-thermodynamic quantities with respect to the TMP by measuring the state of the total system, involving probability distributions for the subsystems at the initial time and the final time. In addition, the quantum mechanical mutual information content is introduced, manipulating both the total and the subsystems. The statistical average of the information-thermodynamic quantities leads us to a new thermodynamic inequality for heat transferred from a heat reservoir, which is bound by the entropy change of subsystems and the change of the quantum mutual information between the subsystems. We discuss our results in an instance of quantum computing and in accordance with the thermodynamic relations obtained elsewhere in the literature.

We start by considering a non-equilibrium process for a bipartite system $\rho_{AB} \in \mathcal{H}_A \otimes \mathcal{H}_B $ comprised of two subsystems, $A$ and $B$ with $\rho_A \in \mathcal{H}_A$ and $\rho_B \in \mathcal{H}_B$, and a reservoir $\rho_R$. An arbitrary quantum bipartite state $\rho_{AB}^i$ initially decoupled from $\rho_R^i$ evolves into a final state $\rho_{ABR}^f = U \rho_{AB}^i \otimes \rho_R^i U^{\dagger}$, where the process is described by a unitary operator $U$. We assume the initial state of $R$ to be a thermal equilibrium state $\rho_R^i = e^{-\beta H_R}/Z_{\beta}$ with an inverse temperature $\beta$ and the corresponding partition function $Z_{\beta}$, where $H_R$ is the Hamiltonian of $R$. The density operators of each subsystem are given by the partial trace of the bipartite state, i.e, $\rho_{A(B)} = \tr_{B(A)}(\rho_{AB})$. The spectral decompositions of $AB$, $A$, $B$, $R$ at the initial time $t_i$ are denoted as $\rho_{AB}^{i}  = \sum_{m} p_m \ket{m} \bra{m}$, $\rho_{A}^{i}  = \sum_{a} p_a \ket{a} \bra{a}$, $\rho_{B}^{i}= \sum_b p_b\ket{b} \bra{b}$, and $\rho_R^i= \sum_{r} p_r\ket{r} \bra{r}$, respectively. In addition, at the final time $t_f$, the spectral decomposition of the states are given by $\rho_{AB}^{f}  = \tr_R(\rho_{ABR}^{f}) = \sum_{m'} p_{m'}\ket{m'} \bra{m'}$,$\rho_{A}^{f}  = \sum_{a'} p_{a'}\ket{a'} \bra{a'}$, $\rho_{B}^{f}= \sum_{b'} p_{b'}\ket{b'} \bra{b'}$. 

Given a joint probability of two random variables, the classical stochastic mutual information $J$  can be defined as  $J(k,l) := \ln [p_{k,l}/p_{k}p_{l}]$, where $p_k$ and $p_l$ are the marginal probabilities of the joint probability $p_{k,l}$. We define the joint probability as $p_{a,b} := \bra{b}\bra{a}\rho_{AB}^i\ket{b}\ket{a}$ at $t_i$ and $p_{a,'b'} := \bra{b'}\bra{a'}\rho_{AB}^f\ket{b'}\ket{a'}$ at $t_f$ for subsystems $A$ and $B$ respectively. Hence, the initial and final classical mutual information are given by $J_i := \ln [p_{a,b}/p_{a}p_{b}]$ and $J_f := \ln [p_{a',b'}/p_{a'}p_{b'}]$, respectively \cite{TS12}. The average of $J$ is equal to the classical mutual information $\sum_{k,l} p_{k,l} J(k,l) = H(A) + H(B) + H(A,B)$, where $H(A)$ is the Shannon entropy of system $A$, $H(B)$ is the Shannon entropy of system $B$, and $H(A,B)$ is the Shannon joint entropy of systems $A$ and $B$.  Note that the classical mutual information represents only the classical correlation between the two subsystems, although the total system is set to be a quantum correlated state. Thus, the fluctuation theorem for quantum bipartite systems requires an information-thermodynamic quantity that describes the total correlations between the subsystems in arbitrary quantum bipartite systems.

In order to resolve this issue, we introduce a mutual information content $I$. Instead of the joint probability $p_{k,l}$, the probability $p_{s}$ of the total system is exploited to define $I$ as $I(s,k,l) :=  {\rm ln} [p_{s}/p_{k}p_{l}]$, where $\rho_{AB} = \sum_{s} p_{s}\ket{s} \bra{s}$, $\rho_{A} = \sum_{k} p_{k}\ket{k} \bra{k}$, and $\rho_{B} = \sum_{l} p_{l}\ket{l} \bra{l}$. Comparing it to the classical stochastic mutual information $J$, it holds one more index, $s$, which is represented by the state of the total system. To find the average mutual information content, the joint probability that involves three indices including $s$ for the total system and $k,l$ for the subsystems is required. Since the projectors of the total system $\Pi_{s} := \ket{s}\bra{s}$ in general do not commute with the products of the projectors of the subsystems $\Pi_{k} := \ket{k}\bra{k}$ and $\Pi_{l} := \ket{l}\bra{l}$, we are not allowed to define the probability that the total system is found in the state $\ket{s}$ and, at the same time, that each subsystem is found in the state $\ket{k}$ for $A$ and $\ket{l}$ for $B$. Hence, using the conditional probability that indicates the subsystem to be found in $\ket{k,l} := \ket{k} \otimes \ket{l}$ provided that the total system to be found in $\ket{s}$, the probability is defined by 
\begin{equation}
p_{s,k,l} = p_s p_{k,l|s} :=\bra{s}\rho_{AB}^i\ket{s}|\langle s | k, l \rangle|^2, \label{jp}
\end{equation}
where $p_s = \bra{s}\rho_{AB}^i\ket{s}$ and $p_{k,l|s} = |\langle s | k, l \rangle|^2$. Similarly, we denote the joint probability of the consecutive measurement at the final time as $p_{s',k',l'} := \bra{s'}\rho_{AB}^f\ket{s'}|\langle s' | k', l' \rangle|^2$ \cite{JK33,YT37,Dirac45,Barut57,Margenau61}. The joint probability yields the average of $I$ which is given by $\langle I \rangle = - \sum_{s,k,l} p_{s,k,l} {\rm ln}[p_{s}/p_{k}p_{l}] = S(\rho_A^i) + S(\rho_B^i) -S(\rho_{AB}^i)$, where $S(\rho) = -\tr(\rho \ln \rho)$ is the von Neumann entropy. $\langle I \rangle$ is the quantum mutual information which is a measure of the total correlation of a quantum bipartite system \cite{NC00}. Note that, in general, $p_sp_{k,l|s} \neq p_{k,l}p_{s|k,l}$,  where $p_{k,l}p_{s|k,l} :=\bra{k,l}\rho_{AB}^i\ket{k,l}|\langle k, l | s \rangle|^2$. In addition, the information contents at $t_i$ and $t_f$ are given by $I_i ={\rm ln} [p_{m}/p_{a}p_{b}]$ and $I_f :={\rm ln} [p_{m'}/p_{a'}p_{b'}]$, respectively.

As we pointed out, the joint probability evaluated by the conventional TMP has its drawbacks \cite{AA14,PT16}. So in order to include all the informational-thermodynamic quantities in the fluctuation theorem, we introduce a joint probability by manipulating the TMP based on the measurements of the total system, including the probability distributions for the subsystems at $t_i$ and $t_f$. The probability for the system $AB$ and the reservoir $R$ to be found in $\ket{m}$ and $\ket{r}$ at $t_i$, and $\ket{m'}$ and $\ket{r'}$ at $t_f$ is defined as 
\begin{equation}
p_{m,m';r,r'} = |\bra{m',r'} U \ket{m,r}|^2 p_m p_r.
\end{equation}
We then multiply the conditional probabilities $p_{a,b|m}$ and $p_{a',b'|m'}$ by $p_{m,m';r,r'}$ to define the probability that indicates the subsystem $A$ and $B$ to be found in the states $\ket{a}$ and $\ket{b}$ provided that the system $AB$ and the reservoir $R$ to be found in the states $\ket{m}$ and $\ket{r}$ at $t_i$, and the subsystem $A$ and $B$ to be found in the states $\ket{a'}$ and $\ket{b'}$ provided that the system $AB$ and the reservoir $R$ to be found in the states $\ket{m'}$ and $\ket{r'}$ at $t_f$  as 
\begin{equation}
p_{m,a,b,m',a',b';r,r'} = p_{m,m';r,r'}|\langle m | a, b \rangle|^2 |\langle m' | a', b' \rangle|^2. 
\end{equation}
 We would like to emphasize that the total system is measured in the TMP, instead of measuring the subsystems directly. To see the validity of the probability, we sum $p_{m,a,b,m',a',b';r,r'}$ over all $m',a',b',r'$. Then, we find $\sum_{m',a',b',r'} p_{m,a,b,m',a',b';r,r'} = p_{m,a,b;r} = |\langle m | a, b \rangle|^2 p_mp_r$. In the same way, we can verify the marginal probabilities $p_{m',a',b';r'}$, $p_m$, etc. Note that, as in all other time-local approaches, we may also consider the multiple point measurement with infinitesimal time steps. Our definition can be straightforwardly extended to $P_{m_0,m_1,...,r_0,r_1,...}|\langle m_0 | a_0, b_0 \rangle|^2 |\langle m_1 | a_1, b_1 \rangle|^2 ...$

As a final step before discussing our main result, consider a joint probability in a time reversed manner. A unitary operator $\tilde{U} = \Theta U \Theta$ describes the time reversed process corresponding to $U$, where $\Theta$ is a time reversal operator. Also, the initial state of the bipartite system and the reservoir in the time reversed process is given by $\tilde{\rho}_{AB}^i \otimes \tilde{\rho}_{R}^i$, where $\tilde{\rho}_{R}^i = e^{-\beta H_R}/Z_{\beta} = \sum_{r'} \tilde{p}_{r'}\ket{r'} \bra{r'}$. Note that the initial state of the total system in the time reversed process is the same state as the final state of the total system in the time forward process, i.e., $\tilde{\rho}_{AB}^i = \rho_{AB}^f$. The corresponding joint probability to the time reversed process is then given by 
\begin{equation}
\tilde{p}_{m,a,b,m',a',b'} =\tilde{p}_{m',m;r',r}  |\langle m' | a', b' \rangle|^2 |\langle m | a, b \rangle|^2,
\end{equation}
where $\tilde{p}_{m',m;r',r} = |\bra{m,r} \tilde{U} \ket{m',r'}|^2 \tilde{p}_{m'}\tilde{p}_{r'}$. This probability will be used to show the Crooks relation later.

With arbitrary initial bipartite states, our main result is
\begin{equation}
\langle e^{- \Delta s_A - \Delta s_B + \Delta I + \beta Q}\rangle = 1, \label{main}
\end{equation}
which takes into account fluctuations in the entropy of subsystems, heat, and quantum mechanical mutual information content that measures correlations between the subsystems. We define the change in the stochastic entropy of the subsystems, the heat transferred from $R$ to $AB$, and the change of the information content as $\Delta s_A : = -\ln \tilde{p}_{a'} - (-\ln p_{a})$, $\Delta s_B : = -\ln \tilde{p}_{b'} - (-\ln p_{b})$, $\beta Q := \ln \tilde{p}_{r'} - \ln p_{r} = \beta[E_{r}^R - E_{r'}^R]$, where $p_r = e^{-\beta E_r^R}/Z_{\beta}$ and $\tilde{p}_{r'} =e^{-\beta E_{r'}^R}/Z_{\beta}$, and $\Delta I := I_f -I_i$, respectively. The main result above will be proved later.
 
Using Jensen's inequality we immediately obtain from Eq. \eqref{main}
 \begin{equation}
 \beta \langle Q \rangle \leq  \langle \Delta s_A \rangle + \langle \Delta s_B \rangle - \langle \Delta I \rangle \label{ineq}
 \end{equation}
 which is our thermodynamic inequality. The statistical averages of the stochastic entropy change of the subsystems and the change of the quantum mechanical mutual information content are given by $\langle \Delta s \rangle = S(\rho_f) - S(\rho_i)$ and $\langle \Delta I \rangle = S(A:B)_f - S(A:B)_i$, where $S(A:B) = S(\rho_A) + S(\rho_B) -S(\rho_{AB})$ is the quantum mutual information \cite{NC00}. Inequality \eqref{ineq} describing the total heat transferred from a reservoir is bound by  the entropy change of individual systems and their correlations in quantum bipartite systems. Therefore, inequality \eqref{ineq} explains the fundamental limit for heat transfer from a thermal reservoir to a quantum bipartite system.

We illustrate inequality \eqref{ineq} using an example from quantum computing. Let us consider a NAND gate as an irreversible logical process which produces the correlation between two bits of inputs and a bit of output. Due to Landauer's principle, the heat is expected to be dissipated in the process, since the size of information required for the input states is larger than that of information for the output states, which leads to the increase of the entropy of environment. Note, however, that the process of NAND gate can be replaced by the process in the reversible computing to avoid the loss of the information.

We consider the Toffoli gate in the reversible computation. To this end, the input states and the output states are embedded into the states of three bits, denoting the initial two bits as the input state $\rho_A$ and the final bit as the output state $\rho_B$. We prepare a superposed input state with two bits of memories, i.e., $\rho_A^i = \ket{\Psi}\bra{\Psi}$, where $\ket{\Psi} = \frac{1}{2}\ket{00} + \frac{1}{2}\ket{01} + \frac{1}{2}\ket{10} + \frac{1}{2}\ket{11} = \frac{1}{\sqrt{2}} (\ket{0} + \ket{1}) \otimes \frac{1}{\sqrt{2}} (\ket{0} + \ket{1})$, and a bit of memory in a pure state, $\rho_B^i = \ket{0}\bra{0}$, which is turned out to be the output state. After the Toffoli gate is operated, the total state at the final time is shown to be $\rho_f = \ket{\Phi}\bra{\Phi}$, where $\ket{\Phi} = \frac{1}{2}\left( \ket{00} + \ket{01} + \ket{10} \right) \ket{0} + \frac{1}{2}\ket{11} \ket{1}$, since the Toffoli gate maps the states $ \ket{000}, \ket{010}, \ket{100},$ and  $\ket{110}$ into $\ket{000}, \ket{010}, \ket{100},$ and $\ket{111}$, respectively. Note that the process of quantum computing starts in the product state with the subsystems superposed, and ends up in an entangled state between inputs and outputs.

In the above process, it indicates that the irreversibility occurs in the subsystems since the entropy of each subsystem increases by $\ln 2$. The state of the total system is, however, transformed from a pure state to a pure state so that the process as seen by the total system is of course reversible. Indeed, inequality \eqref{ineq} shows that the key to reconciling the gap between the irreversibility of the subsystems and the reversibility of the total system lies in the total correlations between the subsystems. Since it is a reversible computation, the gate operates in absence of any heat dissipation, indicating no interaction with any heat reservoir during the process. The average transferred heat during the process is equal to zero, and the changes in the entropy of the subsystems and the quantum mutual information are given by $\langle \Delta s_{A,B} \rangle = \ln 2$ and $\langle \Delta I \rangle = 2\ln 2$, which satisfies inequality \eqref{ineq}.

We next discuss the fluctuation theorem derived by the conventional TMP and show them to be special cases of our result. Assuming the eigenbasis of the joint state of the initial and final state are represented as the product state of eigenbasis of the subsystems, i.e., $\ket{m}=\ket{a''}\otimes\ket{b''}=\ket{a'',b''}$ and $\ket{m'}=\ket{a'''}\otimes\ket{b'''}=\ket{a''',b'''}$, it is shown that $p_{m,a,b,m',a',b';r,r'} = p_{a,b,a',b';r,r'}$, since $|\langle m | a, b \rangle|^2 = |\langle a'', b'' | a, b \rangle|^2 = \delta_{a,a''}\delta_{b,b''}$ and $|\langle m' | a', b' \rangle|^2 = |\langle a''',b''' | a', b' \rangle|^2 = \delta_{a',a'''}\delta_{b',b'''}$. The resulting probability $p_{a,b,a',b';r,r'}$ is equal to the probability obtained by the TMP with the eigenbasis of the subsystems. As a result, Eq. \eqref{main} reduces to 
\begin{equation}
\langle e^{- \Delta s_A - \Delta s_B + \Delta J + \beta Q}\rangle = 1 \label{classical},
\end{equation}
where $\Delta J = J_f - J_i$ is the change in the unaveraged classical mutual information and the statistical average is performed by $p_{a,b,a',b';r,r'}$. Equation \eqref{classical} shows the fluctuation theorems for bipartite systems when the joint state does not involve quantum correlations \cite{SJ15,MM11,KF13}.

Consider a classical observer of subsystem $B$ and a classical memory of $A$. The state of $B$ is assumed to be unchanged and no energy is allowed to be exchanged with external systems. Equation \eqref{classical}, then, reproduces the main result in \cite{TS12}, $\langle e^{- \sigma + \Delta J}\rangle = 1$,  which leads to 
\begin{equation}
\beta \langle Q \rangle \leq \langle \Delta s_A \rangle  - \langle \Delta J\rangle, \label{cLandauer}
\end{equation} where  $\sigma:=\Delta s_A -\beta Q$. Moreover, inequality \eqref{cLandauer} can be interpreted as Landauer's principle in the presence of classical correlations between a memory and an observer. In our quantum mechanical setting, inequality \eqref{ineq} becomes
\begin{equation}
\beta \langle Q \rangle \leq \langle \Delta s_A \rangle  - \langle \Delta I\rangle, \label{qLandauer}
\end{equation} 
when the state of the observer $B$ is assumed to be degenerate and unchanged, i.e., $\langle \Delta s_B \rangle = 0$. Inequality \eqref{qLandauer} indicates the quantum mechanical bound for the heat dissipation can reach negative values during the process of erasure of memory if the memory is quantum-mechanically correlated with an observer. This implies the environment can be cooled in the course of erasure of memory in accordance with the result in Ref. \cite{JJ13, LR11}. 

We now turn to the proof of the main result along with inequality \eqref{ineq}. Given the joint probability for the forward process and the backward process, the Crooks theorem is shown by noting 
\begin{eqnarray}
& &p_{m,a,b,m',a',b';r,r'}\nonumber\\ 
&=& |\bra{m',r'} U \ket{m,r}|^2|\langle m | a, b \rangle|^2 |\langle m' | a', b' \rangle|^2p_{m}p_{r}\nonumber\\
&=& |\bra{m,r} \tilde{U} \ket{m',r'}|^2 |\langle m | a, b \rangle|^2 |\langle m' | a', b' \rangle|^2\tilde{p}_{m'}\tilde{p}_{r'}\frac{p_{m}p_{r}}{\tilde{p}_{m'}\tilde{p}_{r'}}\nonumber\\
&=& \tilde{p}_{m',a',b',m,a,b;r',r}\frac{p_{a}}{\tilde{p}_{a'}}\frac{p_{b}}{\tilde{p}_{b'}}\frac{p_{m}}{p_{a}p_{b}}\frac{\tilde{p}_{a'}\tilde{p}_{b'}}{\tilde{p}_{m'}} \frac{p_{r}}{\tilde{p}_{r'}}\nonumber\\
&=& \tilde{p}_{m',a',b',m,a,b;r',r}e^{\Delta s_A}e^{\Delta s_B}e^{- \Delta I }e^{-\beta Q},\nonumber\\
\end{eqnarray}
where the third line follows from $|\bra{m',r'} U \ket{m,r}|^2 = |\bra{m,r} \tilde{U} \ket{m',r'}|^2$ \cite{TS13} and the fifth line follows from the definition of $\Delta s_A :=\ln \frac{p_{a}}{\tilde{p}_{a'}},  \Delta s_B := \ln \frac{p_{b}}{\tilde{p}_{b'}}, \Delta I := \ln \frac{\tilde{p}_{a'}\tilde{p}_{b'}}{\tilde{p}_{m'}} - \ln \frac{p_{m}}{p_{a}p_{b}}$, and $\beta Q:= - \ln \frac{p_{r}}{\tilde{p}_{r'}}$. The detailed fluctuation theorem for a bipartite system is, then, given by
\begin{equation}
\frac{\tilde{p}_{m',a',b',m,a,b;r',r}}{p_{m,a,b,m',a',b';r,r'}}  = e^{- \Delta s_A - \Delta s_B + \Delta I + \beta Q}.\label{crooks}
\end{equation}
In the end, the main result is verified by the statistical average of the Crooks relation,
\begin{eqnarray}
& &  \langle e^{- \Delta s_A - \Delta s_B + \Delta I + \beta Q}\rangle  \nonumber\\
&=& \sum p_{m,a,b,m',a',b';r,r'}\frac{\tilde{p}_{m',a',b',m,a,b;r',r}}{p_{m,a,b,m',a',b';r,r'}}\nonumber\\
&=& 1,\nonumber\\
\end{eqnarray}
since $\sum \tilde{p}_{m',a',b',m,a,b} = 1$. In addition, we find that 
\begin{eqnarray}
& &\langle \Delta s_A + \Delta s_B - \Delta I - \beta Q \rangle \nonumber\\
&=& \sum p_{m,a,b,m',a',b';r,r'}\ln \frac{p_{m,a,b,m',a',b';r,r'}}{\tilde{p}_{m',a',b',m,a,b;r',r}}.\label{relative}\nonumber\\
\end{eqnarray}
The inequality \eqref{ineq} is also verified by Eq. \eqref{relative} as a consequence of the positivity of the relative entropy.  

In summary, we have derived an information fluctuation theorem when the initial state is quantum correlated. Our fluctuation theorem holds for arbitrary bipartite systems, due to the use of the quantum mechanical mutual information content $I$ and the statistics obtained by the TMP with consecutive measurements. Our approach then leads to a thermodynamic bound for heat transferred from a reservoir given by the entropy of subsystems and their total correlation at the initial time and the final time. Our result generalizes the previously derived thermodynamic inequalities both in the classical and the non-classical limit.

\textit{Acknowledgments}: J.J.P acknowledges funding from the National Research Foundation, Prime Minister's Office, Singapore and the Ministry of Education, Singapore under the Research Centres of Excellence programme. S.W.K. acknowledges funding from the National Research Foundation of Korea (NRF) grants (No.2016R1A2B4015978, 2016R1A4A1008978).
V.V. acknowledges funding from the John Templeton Foundation,
the National Research Foundation (Singapore), the Ministry of Education (Singapore), the Engineering
and Physical Sciences Research Council (UK), the Leverhulme Trust, the Oxford Martin
School, and Wolfson College, University of Oxford. This research is also supported by the National
Research Foundation, Prime Ministers Office, Singapore under its Competitive Research
Programme (CRP Award No. NRF- CRP14-2014-02) and administered by Centre for Quantum
Technologies, National University of Singapore.

\end{document}